\documentclass[english,aps,manuscript]{revtex4}
\usepackage[T1]{fontenc}
\usepackage[latin9]{inputenc}
\usepackage{array}
\usepackage{booktabs}
\usepackage{graphicx}

\makeatletter

\providecommand{\tabularnewline}{\\}

\@ifundefined{textcolor}{}
{%
 \definecolor{BLACK}{gray}{0}
 \definecolor{WHITE}{gray}{1}
 \definecolor{RED}{rgb}{1,0,0}
 \definecolor{GREEN}{rgb}{0,1,0}
 \definecolor{BLUE}{rgb}{0,0,1}
 \definecolor{CYAN}{cmyk}{1,0,0,0}
 \definecolor{MAGENTA}{cmyk}{0,1,0,0}
 \definecolor{YELLOW}{cmyk}{0,0,1,0}
 }

\makeatother

\usepackage{babel}
\begin{document}

\title{Scientific reasoning abilities of non-science majors in physics-based
courses}

\author{J. Christopher Moore}

\email{moorejc@coastal.edu}

\author{Louis J. Rubbo}

\email{lrubbo@coastal.edu}

\affiliation{Department of Chemistry and Physics, Coastal Carolina University,
Conway SC, 29528}
\begin{abstract}
We have found that non-STEM majors taking either a conceptual physics
or astronomy course at two regional comprehensive institutions score
significantly lower pre-instruction on the Lawson's Classroom Test
of Scientific Reasoning (LCTSR) in comparison to national average
STEM majors. The majority of non-STEM students can be classified as
either concrete operational or transitional reasoners in Piaget's
theory of cognitive development, whereas in the STEM population formal
operational reasoners are far more prevalent. In particular, non-STEM
students demonstrate significant difficulty with proportional and
hypothetico-deductive reasoning. Pre-scores on the LCTSR are correlated
with normalized learning gains on various concept inventories. The
correlation is strongest for content that can be categorized as mostly
theoretical, meaning a lack of directly observable exemplars, and
weakest for content categorized as mostly descriptive, where directly
observable exemplars are abundant. Although the implementation of
research-verified, interactive engagement pedagogy can lead to gains
in content knowledge, significant gains in theoretical content (such
as force and energy) are more difficult with non-STEM students. We
also observe no significant gains on the LCTSR without explicit instruction
in scientific reasoning patterns. These results further demonstrate
that differences in student populations are important when comparing
normalized gains on concept inventories, and the achievement of significant
gains in scientific reasoning requires a re-evaluation of the traditional
approach to physics for non-STEM students.
\end{abstract}

\pacs{01.40.H1, 01.40.Di, 01.40.Fk}

\maketitle

\section{Introduction}

University courses in conceptual physics and astronomy typically serve
as students' terminal science experience. Significant work has gone
into developing research-verified pedagogical methods for pre-service
teachers and the algebra- and calculus-based physics courses typically
populated by natural and physical science majors;\cite{key-1} however,
there is significantly less volume in the literature concerning the
non-science, general education population.\cite{key-3} This is quickly
changing, and large, repeatable gains on concept tests are being reported,
specifically within the astronomy education community.\cite{key-2}

Student scientific reasoning and metacognition are increasingly being
investigated within the physics education community, as well, though
most studies use datasets consisting mostly of science majors.\cite{key-8,key-10}
Bao et. al demonstrate that although US college-level science students
perform poorly with respect to physics content knowledge relative
to their Chinese peers, there is no significant difference between
the two groups with respect to scientific reasoning, and that both
groups demonstrate reasonable preparation.\cite{key-10} It is encouraging
that students self-selecting a science major demonstrate competence
in scientific reasoning at the post secondary level. However, there
is little data on the general education student population, which
will comprise a greater percentage of the professional population.

Since most students enrolled in conceptual physics or astronomy will
never take another formal science course, our student learning objectives
should incorporate broader reasoning skills. Scientific reasoning
and metacognitive development are often required for effective decision-making
and problem solving far outside the typical scientific context.\cite{key-11,key-13,key-14}
Furthermore, it has been shown that gains in content knowledge are
strongly correlated to scientific reasoning.\cite{key-8} In particular,
reasoning and meta-cognition development are essential for problem
solving, understanding and applying abstract concepts, and shifting
between multiple representations.\cite{key-3,key-6,key-4} However,
non-STEM majors may enter the classroom with a disadvantage not necessarily
shared by their self-selecting science major peers. Acknowledgement
of the potential dramatic difference in reasoning ability is important
for development of good pedagogy. Furthermore gains in content knowledge
achieved via research-verified, active-engagement curriculum may not
necessarily lead to gains in scientific reasoning.\cite{key-15} In
fact, the content-specific education literature in other disciplines
suggests that explicit intervention is necessary to improve reasoning.\cite{key-16,key-17,key-4,key-3}
It is this explicit intervention that is currently lacking in many
pedagogical models that address this student population in physics
and astronomy, specifically those models practical for implementation
with a large student-to-faculty ratio.

In this paper, we evaluate the post-secondary scientific reasoning
abilities of non-STEM students using Lawson's Classroom Test of Scientific
Reasoning (LCTSR).\cite{key-36,key-24} The purpose is to determine
where the general education population is with respect to scientific
reasoning in order to inform future pedagogies aimed at improving
scientific reasoning within this group. We classify students into
one of three formal reasoning levels as described by Piaget's theory
of cognitive development: concrete operational, transitional, and
formal operational.\cite{key-27} Furthermore, we investigate the
types of content with which non-STEM students struggle, and correlate
learning gains for this content to preparation in scientific reasoning.
Finally, we look at the effectiveness of {}``reformed'' pedagogy
for learning gains in scientific reasoning, and discuss possible implications
for instruction.

\section{Background}

In this section, we review the Piagetian levels of formal reasoning,
the assessment of scientific reasoning, and the connection between
reasoning level and potential gains in specific types of content knowledge.
We also discuss our specific student population and the types of courses
in which these students enroll.

What exactly constitutes scientific reasoning is both complex and
debatable. Lawson suggests that scientific reasoning has a structure
that is chiefly hypothetico-deductive in nature and consisting of
interrelated aspects, such as proportional reasoning, control of variables,
probability reasoning and correlation reasoning.\cite{key-18,key-19}
Inductive and deductive process are involved, with some researchers
intimately linking reasoning with the process of drawing inferences
from initial premises.\cite{key-14,key-20}

More recently, Kuhn has suggested that scientific reasoning is more
than inductive inference, but a truth-seeking social process that
involves the coordination of theory and evidence.\cite{key-22} Kuhn
and others specifically suggest that reasoning process cannot be separated
from prior knowledge.\cite{key-22,key-23,key-24,key-25} Similarily,
the learning of content and reasoning development have been linked
in the physics education literature.\cite{key-8,key-26} As we will
discuss further, content gains are significantly more difficult to
achieve with underprepared students vs. well prepared students, and
gains in reasoning only materialize with explicit intervention.

\subsection{Scientific reasoning and concept construction}

Piaget's theory of cognitive development includes classification into
two formal reasoning levels (concrete operational and formal operational)
with a transitional stage between the two.\cite{key-27,key-32,key-33}
Student's classified as mostly concrete operational reasoners are
characterized by their appropriate use of logic; however, they struggle
with solving problems outside of a concrete context, demonstrating
significant difficulty with abstract concepts and hypothetical tasks.
Formal operational reasoners begin to think abstractly, reason logically,
and draw conclusions from available information. Furthermore, unlike
the concrete operational reasoner, they are able to apply appropriate
logic to hypothetical situations in most contexts. In this way, formal
operational reasoners can begin to think like a scientist, and specifically
develop strong hypothetico-deductive reasoning. Transitional reasoners
fall between the other two classifications where they find success
with hypothetical tasks in some contexts. Lawson describes these levels
as Level 0, Low Level 1, and High Level 1, respectively.\cite{key-24,key-34}
Lawson further describes a post-formal level of reasoning, which is
beyond the scope of this study. In this paper, we will use the traditional
Piagetian labels.

It has been shown that the LCTSR can be used as an assessment of formal
reasoning level, and its validity has been established.\cite{key-17,key-18,key-19,key-24,key-36,key-34}
Specifically, Lawson investigated the development of scientific reasoning
and formal reasoning level in introductory college biology.\cite{key-34,key-24}
For physics, Ates et. al correlate formal reasoning level to conceptual
understanding and problem-solving skills in introductory mechanics.\cite{key-26}
In both of these studies, student populations consist primarily of
science majors or science education prospective teachers. In the present
study, we assess formal reasoning level of non-science, general education
students at the college level.

We have used the 2000 revised, multiple-choice edition of the LCTSR,
which assesses reasoning patterns such as proportional reasoning,
control of variables, probability reasoning, correlation reasoning
and hypothetico-deductive reasoning.\cite{key-24} The LCTSR consists
of 12 scenarios followed by two questions each assessing 6 different
scientific reasoning patterns. Each reasoning pattern is addressed
by two question pairs. One question in a pair elicits a response requiring
effective use of the pattern, while the second question has the student
describe the reasoning behind the response. Overall there are 12 questions
designed to assess application of a reasoning pattern, and 12 questions
designed to evaluate the student's personal approach to that application.
Requiring students to commit to a specific reason for their answers
prevents them from achieving correct answers for the incorrect reasons.
In order for a student to receive a {}``correct'' score, they must
answer both questions within a scenario correctly.

With respect to specific content, Lawson has shown that content can
be classified into three categories: descriptive, hypothetical, and
theoretical.\cite{key-34} Content categorized as descriptive includes
concepts having directly observable exemplars. Examples include more
concrete content requiring no more than rote memorization, and/or
directly observable content such as the conservation of mass, bulb
brightness in circuits, reflection and refraction in optics, and moon
phases. Content categorized as theoretical includes concepts without
directly observable exemplars. Examples within physics include the
more abstract concepts of force, energy, and vector fields. Hypothetical
content typically involves observable exemplars; however, typically
over a time or space scale that makes direct observation impossible.
Lawson demonstrates that concrete operational reasoners can learn
descriptive content with relative ease, whereas they struggle with
theoretical and hypothetical content. Formal operational reasoners
do relatively well with most content.

\subsection{Student population and course structure}

The dataset used in this study consists of students enrolled in either
a conceptual physics or astronomy course with one of the authors during
the past three years. The conceptual physics course is based on \emph{Physics
by Inquiry} (PbI), which is a guided-inquiry approach to content \textquotedblleft{}in
which the primary emphasis is on discovering rather than memorizing
and in which teaching is by questioning rather than by telling.\textquotedblright{}\cite{key-35}
Over the past three years, this course was taught using a large-enrollment
implementation similar to that reported by Scherr.\cite{key-40} For
two years, this course was taught at Longwood University in Farmville,
VA, which is a primarily-undergraduate, regional comprehensive institution.
For one year this course was taught at Coastal Carolina University
(CCU) in Conway, SC. Coastal Carolina is a similar comprehensive,
regional institution. At Longwood, this course was taught in a traditional
lecture room with between 50-70 students and one instructor. Like
Scherr, students proceeded through PbI materials in groups, but with
whole-class \textquotedblleft{}checkouts\textquotedblright{} rather
than instructor-intensive, individual group checkouts. At CCU, the
course was taught in a lecture room designed for the Student Centered
Activities for Large Enrollment University Physics (SCALE-UP) model.\cite{key-41}
A SCALE-UP course incorporates the high-impact practice of collaborative
assignments and projects by fusing lecture, laboratory and recitation
into a single entity. Although the courses at Longwood were taught
in a traditional classroom setting, the principles of the SCALE-UP
model where implemented as much as was feasible.

The conceptual astronomy course ws taught over the past three years
at CCU in the same SCALE-UP classroom used for the conceptual physics
course. An implementation of \emph{Lecture Tutorials in Introductory
Astronomy} and Peer Instruction was used with this course.\cite{key-42,key-43,key-44}
Typical classes had enrollments of between 30-40 students with two
instructors.

\section{Scientific Reasoning in the non-STEM population}

We have found that students in our conceptual physics and astronomy
courses score significantly lower on the LCTSR compared to students
enrolled in courses typically populated with science majors. Table
\ref{tab:Average-LCTSR-scores} shows average LCTSR pre-instruction
scores (N = 1208, avg. = 75\%) for science and engineering majors
enrolled in a calculus-based introductory physics course, as reported
by Bao, et al.\cite{key-10} The LCTSR was also administered to students
taking a conceptual physics or astronomy course with one of the authors
during the past three years. As shown in table \ref{tab:Average-LCTSR-scores},
this population of students scores significantly lower (N = 109, avg.
= 54\%). We found no significant difference between students in the
conceptual physics and astronomy courses, or between students at the
two institutions.
\begin{table}
\caption{Average LCTSR scores for STEM and non-STEM majors.\label{tab:Average-LCTSR-scores}}

\begin{tabular}{>{\centering}p{1in}>{\centering}p{0.75in}>{\centering}p{0.5in}>{\centering}p{0.5in}}
\toprule 
 & LCTSR \% & N & std. dev.\tabularnewline
\midrule
\midrule 
STEM & 75 & 1208 & 18\tabularnewline
\midrule 
non-STEM & 54 & 109 & 17\tabularnewline
\bottomrule
\end{tabular}
\end{table}

That students in STEM majors demonstrate stronger scientific reasoning
ability is not surprising, since most students typically choose their
major based on their strengths. However, such a dramatic difference
in reasoning ability between STEM and non-STEM students may contribute
to disparities in effectiveness of reformed physics pedagogies. What
works in calculus-based physics courses with natural and physical
science students, may not work in the general-education, conceptual
physics course. 

To better understand our student population, we have investigated
non-STEM student weaknesses with respect to scientific reasoning.
Specifically, we looked at the formal reasoning level for the average
non-STEM student and determined which scientific reasoning patterns
presented the most difficulty for this population.

\subsection{Formal reasoning levels}

Using individual student scores on the LCTSR, we classified students
into three formal reasoning categories: Concrete Operational (CO),
Transitional (T), and Formal Operational (FO). As described by Lawson,
students scoring below 25\% on the LCTSR were classified as operational
reasoners, students scoring between 25\% and 58\% were classified
as transitional reasoners, and students scoring above 58\% were classified
as formal operational reasoners.\cite{key-24} As previously mentioned,
a student must correctly answer both questions within a scenario in
order to receive credit. There are 24 questions on the LCTSR, with
12 different scenarios assessing 6 different reasoning patterns. Percentages
were calculated based on the 12 scenarios.

\begin{figure}
\includegraphics{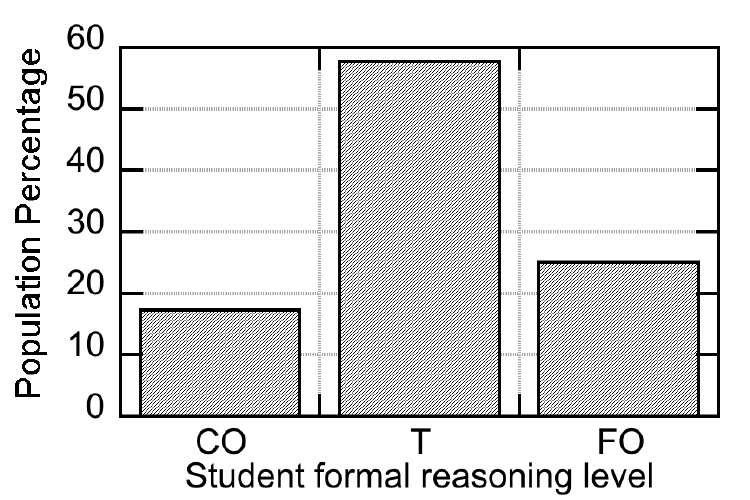}

\caption{Distribution of formal reasoning level for non-STEM students enrolled
in conceptual physics or astronomy. Formal reasoning level was determined
by the LCTSR.\label{fig:Distribution-of-non-STEM}}
\end{figure}

Figure \ref{fig:Distribution-of-non-STEM} shows the distribution
of non-STEM students within Piagetian formal reasoning levels. A significant
majority of non-STEM students (56\%) are classified as transitional
reasoners. This observation is consistent with previous studies of
the general education population in introductory biology courses for
the non-major.\cite{key-45} It should be noted that the majority
of science majors in an introductory, calculus- or algebra-based physics
course would, on average, be classified within the formal operational
category, which is also observed in the biology education literature
for the STEM-based biology introductory courses.\cite{key-46,key-10}

\subsection{Analysis of specific reasoning patterns}

\begin{figure}
\includegraphics{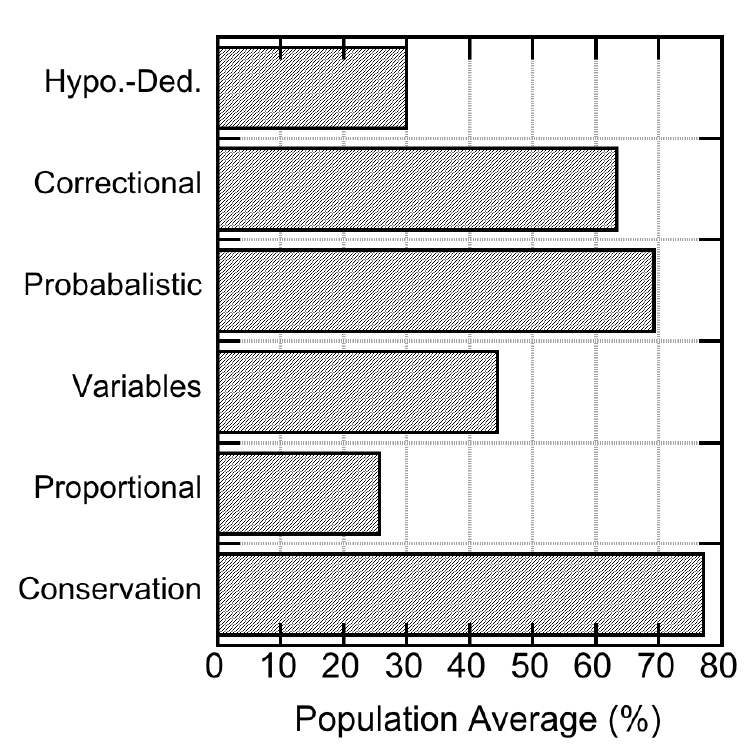}

\caption{Population averages on specific scientific reasoning patterns as assessed
by the LCTSR. Significant difficulty is observed with proportional
reasoning, isolation of variables, and hypothetico-deductive reasoning.\label{fig:LCTSR-patterns}}
\end{figure}

Figure \ref{fig:LCTSR-patterns} shows the population averages for
specific scientific reasoning patterns as assessed by the LCTSR. Students
within the observed population demonstrated significant difficulty
with proportional reasoning, isolation of variables, and hypothetico-deductive
reasoning. Of particular interest, scores on LCTSR questions designed
to test application of hypothetico-deductive reasoning, which can
arguably be called the \textquotedblleft{}scientific method,\textquotedblright{}
average to an abysmal 30\%. Students within this population demonstrated
the poorest performance on proportional reasoning (25\%), which could
be attributable to poor preparation in mathematics. Surprisingly,
most students demonstrate some proficiency with correctional and probabilistic
reasoning. The dramatic difference between performance on probabilistic/correctional
reasoning and proportional reasoning is puzzling. It is possible that
the standard secondary school curriculum includes more explicit instruction
on these reasoning patterns.

As instructors we must be careful not to assume pre-existing knowledge,
specifically with respect to content requiring use of these patterns.
For example, choice of appropriate scale is essential when graphing
data and/or moving between multiple representations. Furthermore,
qualitative approaches to physical systems may be more difficult for
this population, due to a lack of proficiency with proportion. It
is evident that more explicit instruction in scientific reasoning
is necessary, specifically with respect to hypothesis construction
and testing, proportions, and isolation of variables.

\section{Implications for Instruction}

Analysis of students' scientific reasoning level introduces several
implication for instruction. In particular, we expect transitional
and concrete operational reasoners to have difficulty with theoretical
content. This difficulty could result in low learning gains on concept
inventories for these students when compared to formal operational
reasoners receiving similar instruction. As others have previously
demonstrated, pre-existing knowledge and skill sets do contribute
to the maximum achievable learning gain in physics.\cite{key-8}

In this section, we look at the correlation between pre-existing scientific
reasoning ability and learning gains for various classifications of
content. Specifically, is there a stronger correlation between LCTSR
pre-score and normalized learning gain for more theoretical content
in comparison to descriptive content? We also discuss whether content-driven
pedagogies significantly contribute to gains in scientific reasoning.
Is a guided-inquiry based approach to learning science sufficient
for the development of scientific reasoning?

\subsection{Correlation between reasoning and content knowledge gains}

Lawson demonstrated a connection between concept construction and
developmental level in college biology.\cite{key-34} We expect a
similar connection to be prevalent in a physics context. Specifically,
we expect to see a correlation between scientific reasoning and knowledge
gains in a physics-based context. 

\begin{figure}
\includegraphics{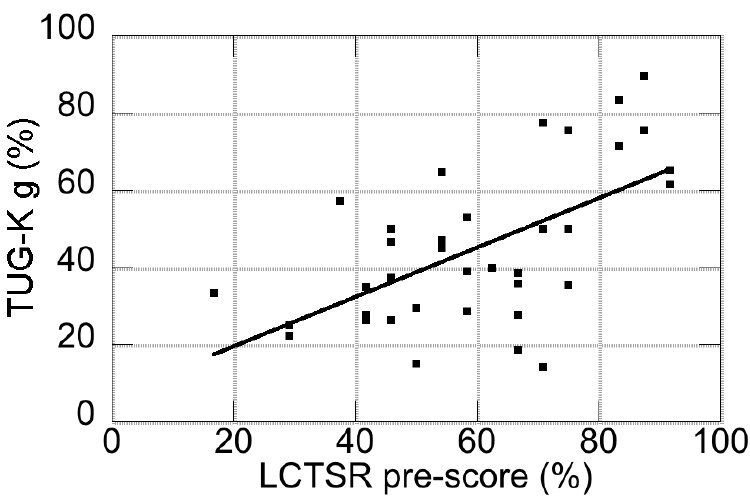}

\caption{Normalized learning gains on TUG-K vs. LCTSR pre-score. The solid
line is the line of best fit (slope=0.64 and r=0.59).\label{fig:TUG-K-vs.-LCTSR}}
\end{figure}

\begin{figure}
\includegraphics{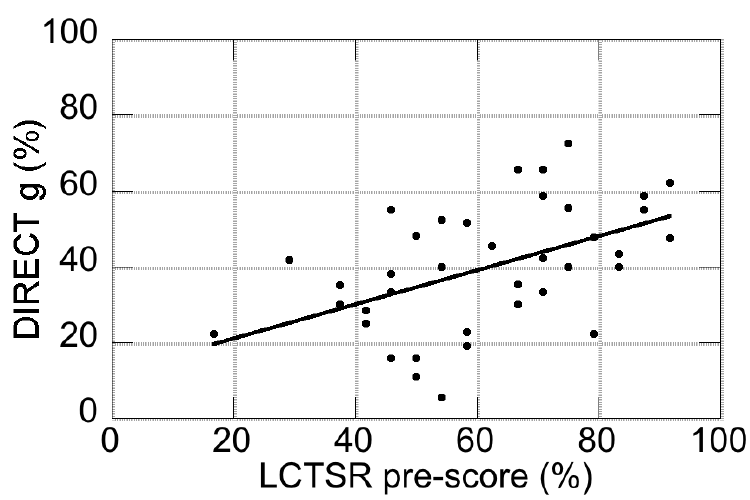}

\caption{Normalized learning gains on DIRECT vs. LCTSR pre-score. The solid
line is the line of best fit (slope=0.45 and r=0.50).\label{fig:DIRECT-vs.-LCTSR}}
\end{figure}

Coletta and Phillips observed a strong correlation between normalized
gain on the FCI and pre-instruction LCTSR scores.\cite{key-8} During
assessment for our conceptual physics courses over the past two years,
we have observed similar strong correlations between pre-instruction
LCTSR scores and normalized gain on two concept inventories, the Determining
and Interpreting Resistive Electric circuits Concept Test (DIRECT)
and the Test for Understanding Graphs -- Kinematics (TUG-K).\cite{key-47,key-48}
As shown in fig. \ref{fig:TUG-K-vs.-LCTSR}, a strong correlation
is seen for content requiring higher-order and more abstract reasoning.
With a slope of linear fit of 0.64 and r=0.59, the correlation between
TUG-K normalized gain and LCTSR score is similar to that seen for
the FCI and stronger than the correlation observed for the DIRECT
assessment {[}slope=0.45 and r=0.50 (see fig. \ref{fig:DIRECT-vs.-LCTSR}){]}.
 
The TUG-K tests a student\textquoteright{}s ability to move between
multiple representations, which rely on higher-order and more abstract
thinking. The FCI assesses a student\textquoteright{}s knowledge and
application of the abstract concept of force. These concepts can be
classified as mostly theoretical, requiring advanced reasoning development
to achieve success. A weaker correlation between DIRECT gains and
LCTSR scores could be because strong scores are possible on DIRECT
via good observation and retention from well-designed inquiry-based
activities; many of the questions are based on content with explicit,
concrete exemplars that are directly observed during the course. This
suggests that if we wish to push our non-science students past the
lower three levels of Bloom\textquoteright{}s Taxonomy of Educational
Objectives, then we may need our courses to focus explicitly on scientific
reasoning early and often.\cite{key-50}

\subsection{Reformed pedagogy and gains in reasoning}

Even with significant disadvantages, substantial gains in content
knowledge can still be obtained in conceptual physics and astronomy
courses, especially when those courses are designed around a research-verified,
active-engagement curriculum. Table \ref{tab:non-explicit-lctsr}
shows average normalized learning gains on DIRECT, TUG-K and the Star
Properties Concept Inventory (SPCI)\cite{key-51} for a subset of
students enrolled in our courses over the past three years. Although
lower than reported for students completing some active-engagement
algebra- and calculus-based courses, these gains are still significant.
Even though we have been relatively successful with content, we have
failed to improve reasoning ability. Average normalized gains on the
LCTSR for both physics and astronomy students are essentially equivalent
to zero. 
\begin{table}
\caption{Average normalized gains on three conceptual inventories and the LCTSR.
Students taking the conceptual physics course were assessed via DIRECT
and TUG-K. Students taking conceptual astronomy were assessed via
the SPCI. Matched data from the LCTSR was obtained from both courses.\label{tab:non-explicit-lctsr}}

\begin{tabular}{>{\centering}p{1in}>{\centering}p{0.5in}>{\centering}p{0.5in}>{\centering}p{0.5in}}
\toprule 
 & g & N & std. dev.\tabularnewline
\midrule
\midrule 
DIRECT & 0.38 & 40 & 0.04\tabularnewline
\midrule 
TUG-K & 0.42 & 38 & 0.05\tabularnewline
\midrule 
SPCI & 0.39 & 36 & 0.05\tabularnewline
\midrule 
LCTSR & 0.06 & 62 & 0.08\tabularnewline
\bottomrule
\end{tabular}
\end{table}

The lack of significant gains in scientific reasoning is particularly
surprising for the conceptual physics course, which via PbI is completely
designed around the process of scientific inquiry. However, it should
be pointed out that we have implemented an adaptation of PbI that
strays in some significant ways from the intentions of the curriculum
designers. Therefore, the data should not be interpreted as condemnation
of any particular pedagogy. Furthermore, this does not to suggest
that gains in reasoning are unachievable. The content-specific education
literature in other disciplines suggests that explicit intervention
is necessary to improve reasoning.\cite{key-16,key-17,key-4,key-3}

\section{Informing Future Pedagogies}

There are three main conclusions from this work that have implications
for physics and astronomy instruction with this population: (1) reformed
pedagogy focused on content alone is not necessarily sufficient to
achieve gains in scientific reasoning; (2) scientific reasoning can
be strongly correlated to gains in content knowledge, especially content
categorized as theoretical; and (3) students entering our conceptual
physics and astronomy courses demonstrate poor preparation in some
scientific reasoning patterns. A dramatic difference in reasoning
ability between STEM and non-STEM students may contribute to disparities
in effectiveness of reformed physics pedagogies. What works in calculus-based
physics courses with natural and physical science students, may not
work in the general-education, conceptual physics course.

In this section we will briefly discuss how these findings can be
used to inform the development of new pedagogies. Specifically, we
discuss the necessity of making scientific reasoning explicit within
our conceptual physics and astronomy courses and the affect topic
sequence could have on student learning.

\subsection{Making scientific reasoning explicit}

The content-specific education literature in other disciplines suggests
that explicit intervention is necessary to improve reasoning.\cite{key-16,key-17,key-4,key-3}
In fact, we are beginning to see significantly larger gains in scientific
reasoning via explicit instruction during our most recent courses,
though these observations are preliminary. 

For example, Lawson presents a series of activities that lead students
through the process of constructing good \textquotedblleft{}if ...
and ... then\textquotedblright{} (IAT) statements in various fields
of knowledge, though with a focus on applications within the biology
content of his courses.\cite{key-17,key-24,key-52} In this way, hypothetico-deductive
reasoning is being made explicit. We have begun introducing these
types of activities within our course with preliminary success. As
an example, PbI has students design an experiment that would address
the question \textquotedblleft{}does a light bulb use up current?\textquotedblright{}
We formalize the process by forcing students to construct an appropriate
IAT statement that is specifically designed to falsify a claim. Based
on observations from a test class consisting of 15 students from both
the physics major and general education population, we have found
that students initially struggle with IAT statement construction but
gradually improve. We combined these activities with activities explicitly
targeting other reasoning patterns (proportional, control of variables,
probability reasoning, correlation reasoning) all mixed within the
actual content as laid out in PbI. 

This explicit approach to reasoning development appears to result
in increased gains on the LCTSR and conceptual inventories. As shown
in tab. \ref{tab:explicit-lctsr}, an average normalized gain of 68\%
(n=14) was achieved on the LCTSR in comparison to 11\% (n=42) for
previous courses taught similarly with respect to content, though
lacking explicit reasoning intervention. Furthermore, content gains
as measured by DIRECT and TUG-K where significantly higher for the
explicit instruction test group, also shown in tab. \ref{tab:explicit-lctsr}.
Although our preliminary sample is small, these preliminary results
are encouraging, and combined with results from the education literature
in other fields, points to a potential high-reward approach. We are
also actively developing activities targeting reasoning patterns for
the conceptual astronomy course.
\begin{table}
\caption{Average normalized gains on two conceptual inventories and the LCTSR
for students participating in a conceptual physics course with and
without explicit reasoning intervention.\label{tab:explicit-lctsr}}

\begin{tabular}{>{\centering}p{1in}>{\centering}p{0.5in}>{\centering}p{0.5in}>{\centering}p{0.5in}}
\toprule 
 & g & N & std. dev.\tabularnewline
\midrule
\midrule 
DIRECT & 0.60 & 15 & 0.04\tabularnewline
\midrule 
TUG-K & 0.55 & 14 & 0.05\tabularnewline
\midrule 
LCTSR & 0.68 & 14 & 0.08\tabularnewline
\bottomrule
\end{tabular}
\end{table}

\subsection{Could topic sequence affect student learning gains?}

Topic sequence could have an effect on student learning gains, even
with explicit instruction in scientific reasoning patterns, especially
with the conceptual physics course. An examination of several conceptual
physics and physical science textbooks shows that the average university
course begins with topics in mechanics, specifically motion, force,
and energy. These more theoretical concepts are the hardest in which
to achieve significant learning gains with this population of students;
however, they are typically the first topics covered in the traditional
conceptual physics course. 

A topic sequence where more concrete content is initially introduced
along with explicit instruction in scientific reasoning may lead to
better performance in theoretical content later in the course. As
an example, much of the content within the topics of circuits and
optics have directly observable exemplars, which makes them ideal
candidates for introductory material. With regards to circuits, McDermott
has shown that no formal instruction in electrostatics or field theories
are needed to successfully build a predictive model.\cite{key-53}
It may be beneficial for students in conceptual astronomy to begin
with observation-based activities and physical model building, such
as the celestial sphere. Gradually increasing the amount of hypothetical
and theoretical content as reasoning develops could build student
confidence in handling physics while better preparing them for theoretical
content.

\section{Summary}

In summary, we have found that non-STEM majors taking either a conceptual
physics or astronomy course at two regional comprehensive institutions
score significantly lower pre-instruction on the LCTSR in comparison
to STEM majors. Furthermore, there is a strong correlation between
pre-instruction LCTSR scores and normalized gains on concept inventories.
The correlation is strongest for content that can be categorized as
mostly theoretical, meaning a lack of directly observable exemplars,
and weakest for content categorized as mostly descriptive, where directly
observable exemplars are abundant. 

Although the implementation of research-verified, interactive engagement
pedagogy can lead to gains in content knowledge, significant gains
in theoretical content (such as force and energy) are more difficult
with this population of students. We also observe no significant gains
on the LCTSR without explicit instruction in scientific reasoning
patterns. This has several implications for instruction, such as the
necessity for explicit instruction, and the potential need for a reassessment
of the canonical sequence of topics in conceptual physics and astronomy.

These results further demonstrate that differences in student populations
are important when comparing normalized gains on concept inventories,
and the achievement of significant gains in scientific reasoning requires
a re-evaluation of the traditional approach to physics for non-STEM
populations.
\begin{acknowledgments}
The authors would like to thank Michelle Parry, Associate Professor
of Physics at Longwood University, for helpful discussions about the
development of the conceptual physics course and scientific reasoning.
We would also like to thank Janelle Bailey, Assistant Professor of
Science Education at the University of Nevada, Las Vegas, for providing
initial analysis of our SPCI raw data.\end{acknowledgments}

\end{document}